\documentclass[twoside,12pt]{article}
\usepackage{epsf,epsfig,amssymb,amsmath,graphicx,float}
\usepackage{color}

\begin{document}
\renewcommand{\thefootnote}{\fnsymbol{footnote}}

\begin{titlepage}

\begin{center}

\vspace{1cm}

{\Large {\bf Relic Density of Asymmetric Dark Matter with Sommerfeld 
enhancement }}

\vspace{1cm}

{\bf Aihemaitijiang Abudurusuli, Hoernisa Iminniyaz}

\vskip 0.15in
{\it
{School of Physics Science and Technology, Xinjiang University, \\
Urumqi 830046, China} \\
}

\abstract{We investigate the evolution of abundance of the asymmetric 
thermal Dark Matter when its annihilation rate
at chemical decoupling is boosted by the Sommerfeld enhancement. 
Then we discuss the effect of kinetic 
decoupling on relic abundance of asymmetric Dark Matter when the 
interaction rate depends on the velocity. Usually the relic 
density of asymmetric Dark Matter is analyzed in the frame of chemical 
decoupling. Indeed after decoupling from the chemical equilibrium, asymmetric 
Dark Matter particles and anti--particles were still in kinetic equilibrium 
for a while. It has no effect on the case of $s-$wave annihilation since 
there is no temperature dependence in this case. However, the kinetic 
decoupling has impacts for the case of $p-$wave annihilation and Sommerfeld 
enhanced $s-$ and $p-$wave annihilations. We investigate in which extent the 
kinetic decoupling affects the relic abundances of asymmetric Dark Matter 
particle and anti--particle in detail. We found the constraints on the cross 
section and asymmetry factor by using the observational data of relic 
density of Dark Matter.  
}
\end{center}
\end{titlepage}
\setcounter{footnote}{0}

\section{Introduction}

There are compelling evidences for the existence of Dark Matter from the
astrophysical and cosmological observations. Despite this evidences, the
nature of Dark Matter is not made clear until now. Asymmetric Dark Matter 
is one of the alternatives which is contrary to the common
assumption that the Majorana particle neutralino could be the candidate for 
Dark Matter which is Weakly Interacting Massive Stable Particles (WIMPs) 
appeared in supersymmetry. The idea for asymmetric Dark Matter 
arises from the possible link between the baryon number density and the
Dark Matter energy density \cite{adm-models,frandsen}. The average density of 
baryons with $\Omega_b = 0.046$ is comparable to that of Dark Matter. It is 
well known that the ordinary matter in the Universe is almost completely
made from baryons, and the anti--baryons are contributing only a small 
fraction. The connection between the baryons and Dark Matter leads to the
assumption that the Dark Matter particles can be asymmetric for which particles 
and anti–particles are not identical and there are more Dark Matter particles 
than anti–particles (or vice versa).  

Refs.\cite{GSV,Iminniyaz:2011yp} discussed the relic abundance of 
asymmetric Dark Matter in the standard cosmological scenario which assumed
the asymmetric Dark Matter particles and anti--particles were in 
thermal equilibrium in the end of 
the radiation dominated era and decoupled when they become
nonrelativistic. In this scenario, usually it is assumed the 
anti--particles are completely annihilated away with their particles and
there are particles in the end. They showed that the final abundances of 
asymmetric Dark Matter particle and anti--particle are determined not only 
by the annihilation cross section, but also by the asymmetry factor which is 
the deviation of co--moving densities of the particle and
anti--particle that is stated later in this paper. 

In this work, we investigate the asymmetric Dark Matter which is coupled to the
sufficiently light force mediators and the interaction between the Dark Matter
particle and anti--particle appeared as long--range interaction. In this case, 
the wavefunction of asymmetric Dark Matter particle and antiparticle is 
distorted by the long--range interaction; it is the Sommerfeld 
effect \cite{Sommerfeld}. The Sommerfeld effect enhances the 
late--time Dark Matter annihilation signals 
\cite{Pospelov:2008jd,MarchRussell:2008tu}. The Sommerfeld enhancement is
determined by the coupling of Dark Matter to the light force mediator. 
Asymmetric Dark Matter needs stronger couplings than the symmetric Dark Matter
of the same mass, then the implications of the Sommerfeld enhancement for the
phenomenology of asymmetric Dark Matter may be quite important than the 
symmetric Dark Matter case.

The effect of Sommerfeld enhancement on the relic density 
for symmetric Dark Matter was already investigated in
\cite{ArkaniHamed:2008qn,Feng:2010zp,Kamionkowski:2008gj,Dent:2009bv,
Zavala:2009mi,Feng:2009hw,Iminniyaz:2010hy}. 
In refs.\cite{Baldes:2017gzw,Agrawal:2017rvu,Feng:2009mn,Petraki:2015hla}, 
the authors discussed asymmetric thermal Dark 
Matter with Sommerfeld enhancement including the effect of the bound state. In 
this paper, we explore the relic density 
of asymmetric Dark Matter particles and anti--particles when the 
annihilation cross section of asymmetric Dark Matter particle 
and anti--particle is enhanced by the Sommerfeld effect. 
Here we only consider the Sommerfeld effect and neglect the effect of 
bound state formation on the relic density of asymmetric Dark Matter.  
We found the particle abundance is not  modified significantly when the 
annihilation rate is boosted by Sommerfeld enhancement. However, for 
Dark Matter anti--particle, the decrease of abundance is more sizable than 
the case of without including the effect of Sommerfeld enhancement. 

Although the asymmetric Dark Matter particles and anti--particles dropped out 
of chemical equilibrium, they were still in kinetic equilibrium for a while 
through the scattering off relativistic standard model particles in the 
thermal plasma. When the annihilating asymmetric Dark Matter particles and 
anti--particles were both in chemical and kinetic equilibrium, the 
temperatures of them tracks the background radiation temperature $T$, i.e. 
$T_{\chi,\bar\chi} = T$. At some point, the rate of scattering falls
below the expansion rate of the universe, then the asymmetric Dark Matter
particles and anti--particles dropped out of kinetic equilibrium. After 
kinetic decoupling, the temperatures of asymmetric Dark Matter particle and 
anti--particle are related by $T_{\chi,\bar\chi} = T^2/T_k$ with the 
background radiation temperature $T$, where $T_k$ is the kinetic decoupling 
temperature \cite{Bringmann:2006mu,Bringmann:2009vf}. The thermal average 
of cross section which is appeared in the Boltzmann equation is different 
before and after kinetic decoupling due to the change of temperature 
dependence. This has impacts on the relic densities of asymmetric Dark 
Matter particles and anti--particles. Without Sommerfeld enhancement, the 
kinetic decoupling has no effect on the relic abundance of asymmetric Dark 
Matter for $s-$wave annihilation since there is no temperature dependency in 
this case. However, there is very small impact in the case of $p-$wave 
annihilation. On the other hand, the effect is more significant both for the 
Sommerfeld enhanced $s-$wave and $p-$wave annihilations. The relic abundance 
of asymmetric Dark Matter is continuously decreased until the Sommerfeld
enhancement ceases to have impact on the relic abundances.

The effect of kinetic decoupling on relic density of Dark Matter for 
the Sommerfeld enhancement was probed in 
refs.\cite{Dent:2009bv,Iminniyaz:2011pva,Hisano:2011dc,Chen:2013bi}.
 Ref.\cite{vandenAarssen:2012ag} discussed the case including 
effect of resonance for $m_{\phi} \neq 0$.  The
impact of early kinetic decoupling on the relic density was also
investigated in ref.\cite{Binder:2017rgn}. In this 
work, we extend this discussion to the asymmetric Dark
Matter. We explore the effects of kinetic decoupling on 
relic abundances of asymmetric Dark Matter particle and anti--particle in 
detail when the annihilation cross section of asymmetric Dark Matter is
changed by Sommerfeld enhancement. Here we discuss the case where the
mediator between asymmetric Dark Matter is massless, $m_{\phi} = 0$.  
We found the relic abundances of asymmetric Dark Matter particle 
and anti--particle are decreased after kinetic decoupling.  
The decrease is almost invisible for asymmetric Dark Matter particle; 
on the other hand, the decrease is sizable for asymmetric Dark Matter 
anti--particle. The magnitude of the decrease depends on the asymmetry factor 
$\eta$, coupling strength $\alpha$ and the kinetic decoupling temperature
$T_k$.

The paper is arranged as following. In section 2, we discuss the thermal
average of the Sommerfeld enhanced annihilation cross section for asymmetric
Dark Matter. In Section 3, we study the numerical solution of
asymmetric Dark Matter abundance including
the effect of Sommerfeld enhancement. The analytic result for the relic
density of asymmetric Dark matter is presented in section 4. In section 5, we 
investigate the effects of kinetic decoupling on 
the relic abundances of asymmetric Dark Matter particle and anti--particle. In 
section 6, the constraints on the parameter space are obtained by using the 
observational data of Dark Matter. In the last section, we summarize our 
results.

\section{Sommerfeld enhanced annihilation cross section}
For a massless light force carrier 
$m_{\phi}$ (in the limit $m_{\phi} \rightarrow 0$), the Sommerfeld factor for 
$s$--wave annihilation is
\begin{equation}\label{eq:Som_s}
S_s = \frac{2 \pi\alpha/v}{1-e^{-2\pi\alpha/v}},  
\end{equation}
and for $p$--wave annihilation
\begin{equation}\label{eq:Som_p}
S_p = 
\left[1 + (\frac{\alpha}{v})^2\right]\frac{2 \pi\alpha/v}{1-e^{-2\pi\alpha/v}},  
\end{equation}
where $v$ is the relative velocity of two
annihilating asymmetric Dark Matter particle and anti--particle,  
$\alpha$ is a coupling strength \cite{Iengo}. Here we only 
consider the annihilation of particle $\chi$ and anti--particle $\bar\chi$. 
When the asymmetric Dark Matter particles and anti–particles decouple from 
the thermal background, they are non–relativistic. Without Sommerfeld 
enhancement, the annihilation cross section for asymmetric Dark Matter 
particle and anti–particle can be expanded with respect to the relative 
velocity $v$,
\begin{equation}\label{eq:cross_a}
     \langle \sigma v \rangle  = a + b\,  \langle v^2 \rangle + 
                       {\cal O}( \langle v^4 \rangle )\, ,  
\end{equation} 
where $a$ is the $s$--wave contribution to $\sigma v$ when $p$--wave is
suppressed, $b$ describes the $p$--wave contribution to $\sigma v$. 
After including Sommerfeld enhancement on the thermal average of annihilation
cross section, we have
\begin{equation}\label{eq:Somcross_a}
     {\langle \sigma v \rangle}_S  = a\,\langle S_s\rangle  + b\,  
               \langle v^2 S_p \rangle+ 
                       {\cal O}( v^4 )\, . 
\end{equation} 
Here we use  $B_s$ and $B_p$ to denote the Sommerfeld boost factors as  
\begin{equation}\label{eq:Somaverage_s}
B_s =  \langle S_s\rangle  = 
      \frac{x^{3/2}}{2 \sqrt{\pi}}\,\int^{\infty}_0 dv~
          v^2 ~ e^{-\frac{x}{4} v^2}\, 
          \frac{2 \pi \alpha/v}{1- e^{- 2 \pi \alpha/v}}\, ,
\end{equation}
and
\begin{equation}\label{eq:Somaverage_p}
B_p =  \langle v^2\, S_p\rangle = 
      \frac{x^{3/2}}{2 \sqrt{\pi}}\,\int^{\infty}_0 dv~
          v^4 ~ e^{-\frac{x}{4} v^2}\, 
          \left[ 1 + (\frac{\alpha}{v})^2 \right]
          \frac{2 \pi \alpha/v}{1- e^{- 2 \pi \alpha/v}}\, .
\end{equation}
Where $x =m/T $ with $m$ being the mass of asymmetric Dark Matter. 
Following we obtain the analytic result of thermal average of Sommerfeld 
enhanced annihilation cross section times relative velocity in approximate way
\cite{Iminniyaz:2010hy}.
For the case, $\pi \alpha /v \ll 1$, we 
expand the factor $(2 \pi \alpha/v)/(1- e^{- 2 \pi \alpha/v})$ in 
Eqs.~(\ref{eq:Somaverage_s},\ref{eq:Somaverage_p}) in Taylor series up to 
the second order, 
\begin{equation}\label{eq:Tay}
      \frac{2 \pi \alpha/v}{1- e^{- 2 \pi \alpha/v}}
       = 1 + \frac{\pi \alpha}{v} + \frac{1}{3} (\frac{\pi \alpha}{v})^2.
\end{equation}
Plugging the Taylor series into 
Eq.(\ref{eq:Somcross_a}), we obtain 
\begin{eqnarray}\label{eq:Tayaverage}
      {\langle \sigma v \rangle}_{S,{\rm Taylor}}   
     & = & a \left( 1 + 
    \alpha \sqrt{\pi x} + \frac{1}{6}\pi^2 \alpha^2 \, x    \right) 
  \nonumber \\
  & + & b \left[\alpha^2 \left( 1 + 
    \alpha \sqrt{\pi x} + \frac{1}{6}\pi^2 \alpha^2 \, x    \right) + 
     \frac{6}{x}~ \left( 1 + \frac{2}{3} \alpha \sqrt{\pi x} + 
        \frac{1}{18}\pi^2 \alpha^2 x \right) \right].
\end{eqnarray}  
When $\alpha=0$, the standard annihilation cross section is recovered. 
In the opposite limit $\pi \alpha/v \gg 1$, $e^{- 2 \pi \alpha/v}$ in the 
denominator of 
Eqs.(\ref{eq:Somaverage_s}) and (\ref{eq:Somaverage_p}) are negligible, then the
cross section is enhanced by $1/v$, we have 
\begin{equation}\label{eq:v_s}
     \langle \sigma v \rangle_{s,{1/v}} = 2 \alpha \sqrt{\pi x},
\end{equation}
\begin{equation}\label{eq:v_p}
     \langle \sigma v \rangle_{p,{1/v}} = 
    8\alpha \sqrt{\pi/x} + 2\alpha^3\sqrt{\pi x}\,.
\end{equation}
Using Eq.(\ref{eq:Tayaverage}) and applying Pade method, we can find the 
well fitting rational functions which connects the two limiting cases and 
can reproduce the exact numerical results for the thermal average of 
annihilation cross section times relative velocity, 
\begin{eqnarray} \label{eq:approx}
      {\langle \sigma v \rangle}_{S,{\rm approx}}  = 
    a\, B_{s, \rm approx} + b\, B_{p, \rm approx},
\end{eqnarray}
where
\begin{eqnarray} \label{eq:approxs}
       B_{s, \rm approx} = 
       \frac{ 1 + 7/4~\alpha \sqrt{\pi x} + 3/2~\alpha^2 \pi x 
       + (3/2 - \pi/3 )~(\alpha^2 \pi x)^{3/2}     }
      { 1 + 3/4~\alpha \sqrt{\pi x} + (3/4 - \pi/6)~ \alpha^2 \pi x },         
\end{eqnarray}
and
\begin{eqnarray} \label{eq:approxp}
      B_{p, \rm approx} & = &
       \alpha^2 \, \frac{ 1 + 7/4~\alpha \sqrt{\pi x} + 3/2~\alpha^2 \pi x 
       + (3/2 - \pi/3 )~(\alpha^2 \pi x)^{3/2}     }
      { 1 + 3/4~\alpha \sqrt{\pi x} + (3/4 - \pi/6)~ \alpha^2 \pi x }
         \nonumber \\
        & + &  \frac{6}{x}\, \frac{ 1+ 4/3~ \alpha \sqrt{\pi x}  
      + (\pi + 4)/9~ \alpha^2 \pi x + 4/51~ \pi~ (\alpha^2 \pi x)^{3/2}  }  
         {1+ 2/3~\alpha \sqrt{\pi x} + \alpha^2 \pi^2  x/18  }\, . 
\end{eqnarray}
We noted that the choice is not unique. The approximation 
reproduces the exact results with accuracy of less than 
$0.5\%$.
%
\begin{figure}[h]
  \begin{center}
    \hspace*{-0.5cm} \scalebox{0.6}{\includegraphics*{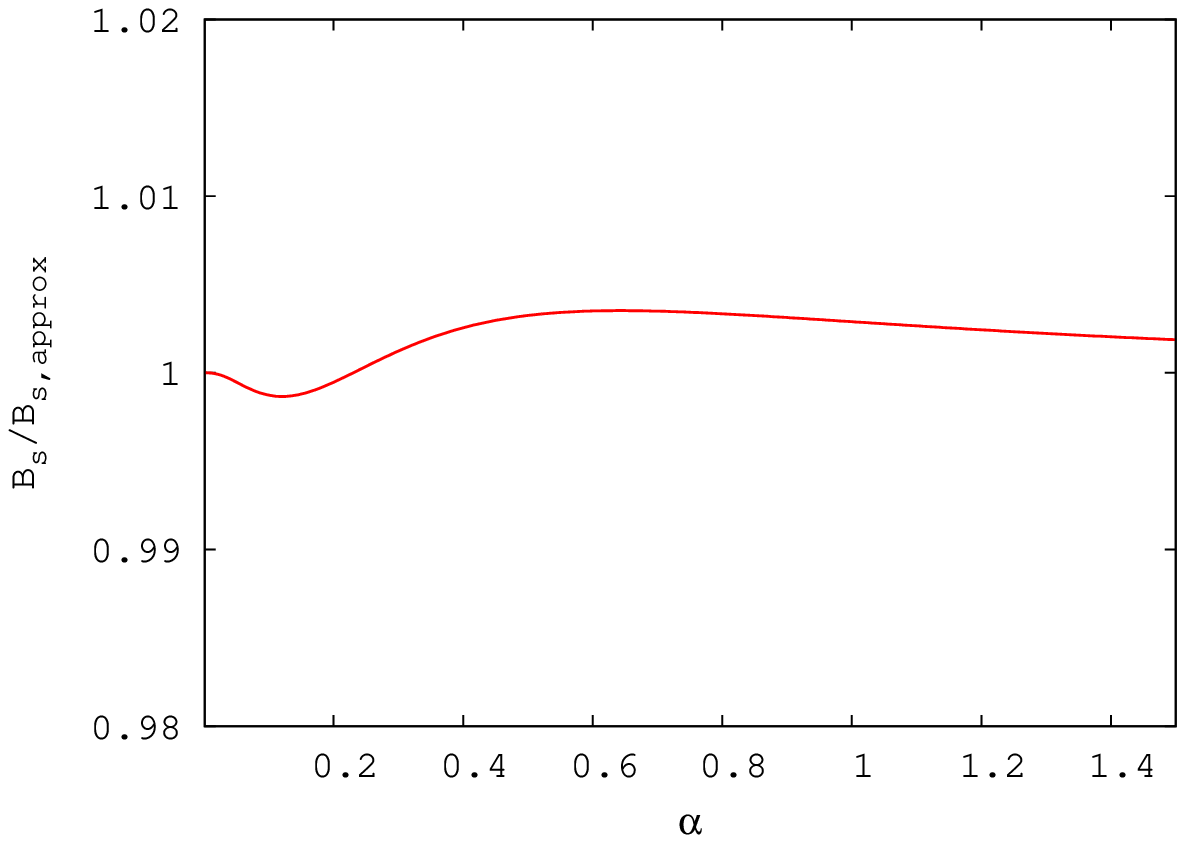}}
      \put(-115,-12){(a)}
            \scalebox{0.6}{\includegraphics*{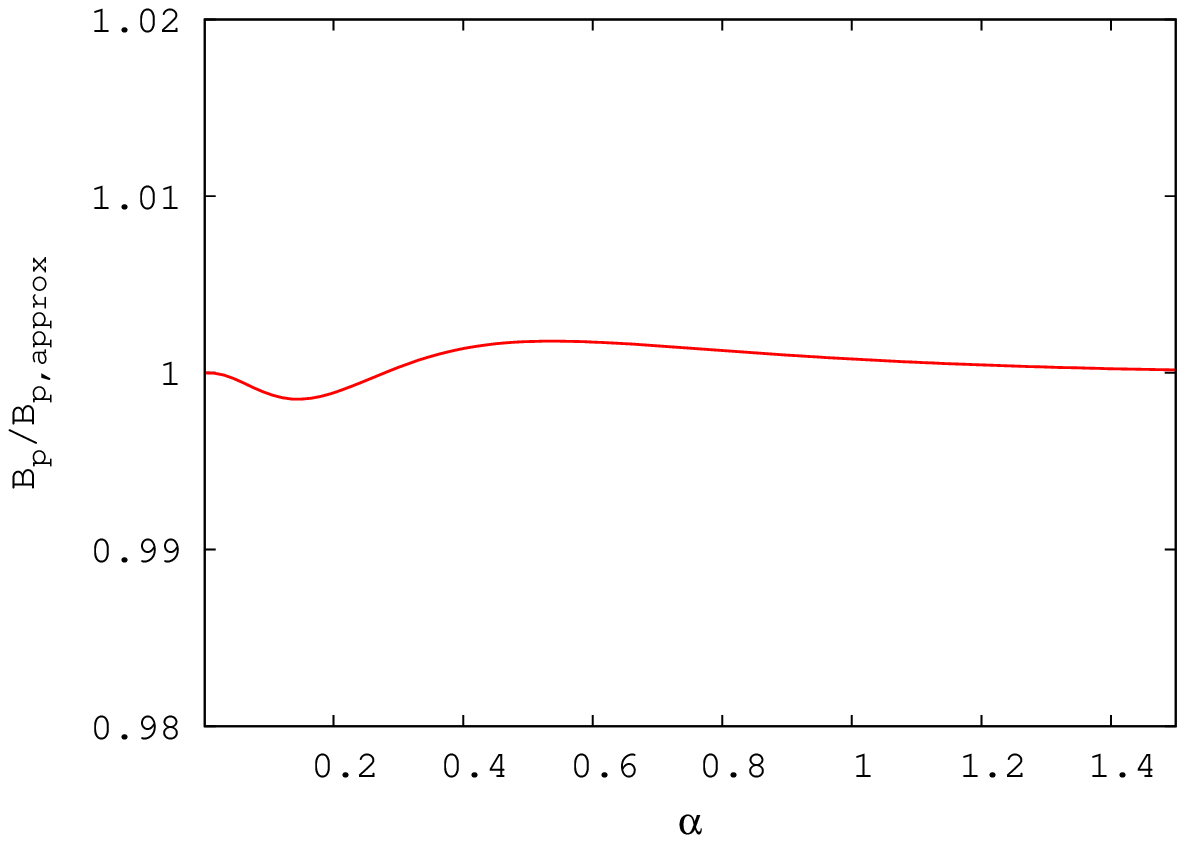}}
    \put(-115,-12){(b)}
       \caption{\footnotesize
       The ratio of the exact value of  
   $B_s $ ($B_p $) and the approximation of 
   $B_{s, \rm approx}$ ($B_{p, \rm approx}$) as a 
   function of $\alpha$ for $m/T = 22$ in $(a)$ ($(b)$). }  
    \label{fig:salpha}
  \end{center}
\end{figure}

Fig.\ref{fig:salpha} shows the ratio of exact values of thermally 
averaged Sommerfeld boost factors 
$B_s$, $B_p$ and the approximation of 
$B_{s, \rm approx}$, $B_{p, \rm approx}$ as a 
function of $\alpha$ for the typical inverse--scaled WIMP decoupling 
temperature $m/T = 22$ in $(a)$ and $(b)$. We found that our 
approximation reproduce the exact results with accuracy of less than 0.5\% 
in both $(a)$ and $(b)$. 
\section{Numerical solution of the abundance of asymmetric Dark Matter
including Sommerfeld enhancement}
After including Sommerfeld enhanced annihilation cross section in the 
Boltzmann equation which describes the evolution of number densities of 
asymmetric Dark Matter particle and anti--particle, we have
\begin{eqnarray} \label{eq:boltzmann_n}
\frac{d n_{\chi,\bar\chi}}{dt} + 3 H n_{\chi,\bar\chi} &=&  
- {\langle \sigma v\rangle}_S (n_{\chi} n_{\bar\chi} - 
n_{\chi,{\rm eq}} n_{\bar\chi,{\rm eq}})\,,
\end{eqnarray}
where $\chi$ is for particle and $\bar\chi$ for anti--particle and the 
expansion rate in the radiation dominated era, 
$H = \pi T^2/M_{\rm Pl} \, \sqrt{g_*/90} $, here 
$M_{\rm Pl} =2.4 \times 10^{18}$ GeV is the reduced Planck
mass, with $g_*$ being the effective number of relativistic degrees of 
freedom. The equilibrium number densities are $n_{\chi,{\rm eq}} = g_\chi 
{\big[m T/(2 \pi) \big]}^{3/2}{\rm e}^{(-m+\mu_\chi)/T}$ and 
$n_{\bar\chi,{\rm eq}} =  g_\chi \,{\big[m T/(2 \pi) \big]}^{3/2}
{\rm e}^{(-m - \mu_{\chi})/T}$.
Here the chemical potentials $\mu_{\chi}$, $\mu_{\bar\chi}$ for $\chi$ and 
$\bar\chi$ are equal when the
asymmetric Dark Matter particle $\chi$ and anti--particle $\bar\chi$ are in 
equilibrium, $\mu_{\bar\chi} = - \mu_{\chi} $, where $g_{\chi}$ is the number
of intrinsic degrees of freedom of the particle.

The Boltzmann equation in terms of the ratio of number 
densities of particle and anti--particle to entropy density 
$Y_{\chi,\bar\chi} = n_{\chi,\bar\chi}/s$, and $x$, is
\begin{equation} \label{eq:boltzmann_Y}
\frac{d Y_{\chi,\bar\chi}}{dx} =
      - \frac{{\lambda \langle \sigma v \rangle}_S}{x^2}\,
     (Y_{\chi}~ Y_{\bar\chi} - Y_{\chi, {\rm eq}}~Y_{\bar\chi, {\rm eq}}   )\,,
\end{equation}
where $s= 2 \pi^2 g_{*s}/45\, T^3$, with $g_{*s}$ being the effective number of 
entropic degrees of freedom. Here we used the entropy conservation, 
$\lambda = 1.32\,m M_{\rm Pl}\, \sqrt{g_*}\,$,
 $g_*\simeq g_{*s}$ and $dg_{*s}/dx\simeq 0$.
The subtraction of the Boltzmann equations for 
$\chi$ and $\bar\chi$ results 
\begin{equation}\label{eq:eta}
      \frac{d Y_{\chi}}{dx} - \frac{d Y_{\bar\chi}}{dx} = 0.
\end{equation}
This means
\begin{equation}\label{eq:eta}
      Y_{\chi} - Y_{\bar\chi} = \eta\,,
\end{equation}
where $\eta$ is a constant, the difference of the co--moving densities of the
particles and anti--particles is conserved. Inserting this into Boltzmann 
equation (\ref{eq:boltzmann_Y}), then
\begin{equation} \label{eq:Yeta}
\frac{d Y_{\chi}}{dx} =
     - \frac{{\lambda \langle \sigma v \rangle}_S}{x^2}~  
     (Y_{\chi}^2 - \eta Y_{\chi}  - Y^2_{\rm eq})\,,
\end{equation}
\begin{equation} \label{eq:Ybareta}
\frac{d Y_{\bar\chi}}{dx} =
     - \frac{{\lambda \langle \sigma v \rangle}_S}{x^2}~  
     (Y_{\bar\chi}^2 + \eta Y_{\bar\chi}  - Y^2_{\rm eq})\,,
\end{equation}
here 
$Y^2_{\rm eq}= Y_{\chi,{\rm eq}} Y_{\bar\chi,{\rm eq}}=(0.145g_{\chi}/g_*)^2\,x^3e^{-2x}$. 
We noted that $Y^2_{\rm eq}$ doesn't depend on the chemical potential 
$\mu_{\chi}$.

In the standard picture of particle evolution scenarios, it is assumed 
the asymmetric Dark Matter particles and anti--particles were in thermal 
equilibrium with the standard model plasma in the early universe. They 
decoupled from equilibrium whenever the interaction rate $\Gamma$ drops below 
the expansion rate $H$. At this point the temperature is less than the mass of 
asymmetric Dark Matter particles, $T < m$ for $m > |\mu_{\chi}|$ 
\cite{GSV,Iminniyaz:2011yp,standard-cos}. This is the freeze out
temperature at which point the number densities of asymmetric Dark Matter 
particle and anti--particle in a co--moving space almost become constant.
\begin{figure}[h]
  \begin{center}
    \hspace*{-0.5cm} \includegraphics*[width=8cm]{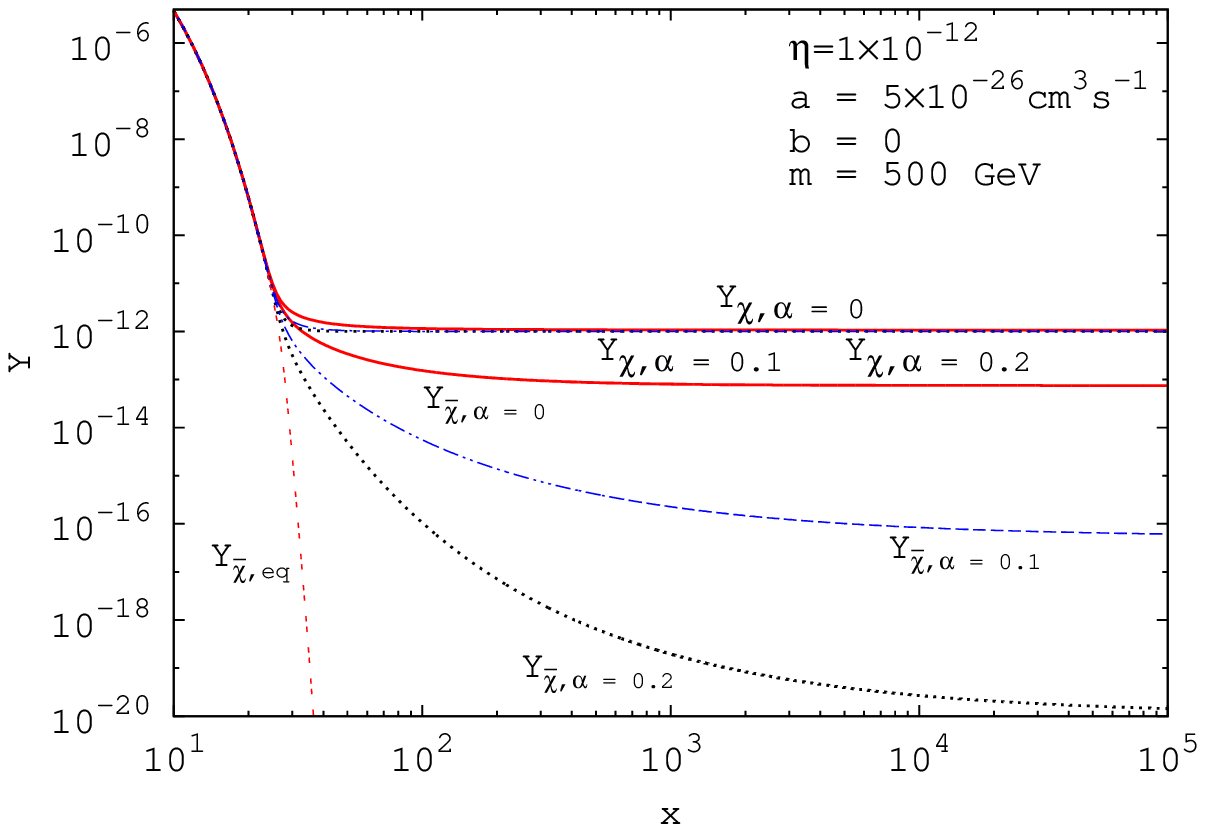}
    \put(-115,-12){(a)}
    \hspace*{-0.5cm} \includegraphics*[width=8cm]{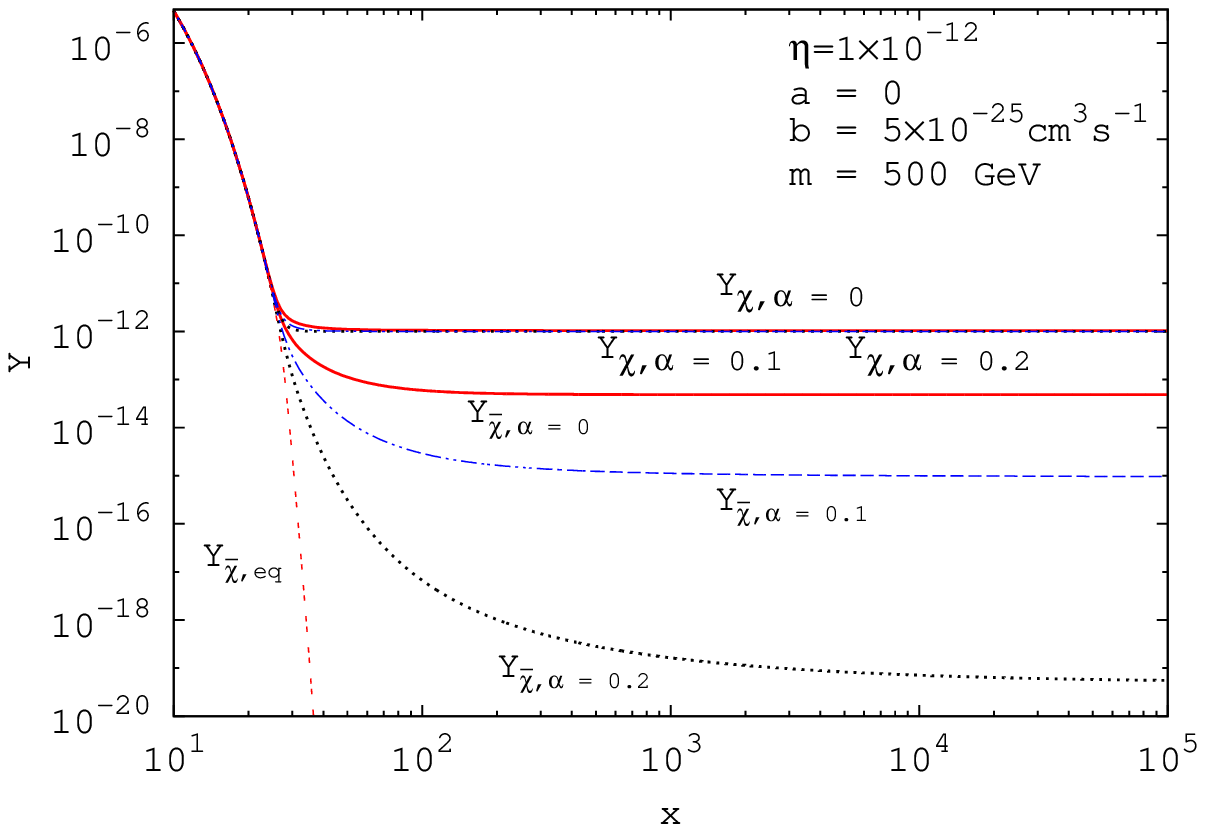}
    \put(-115,-12){(b)}
     \caption{\label{fig:a} \footnotesize Evolution of $Y$ for the particle
       and anti--particle as a function of $x$ for the case when the
       annihilation cross section is boosted by Sommerfeld enhancement and 
       without enhancement. Here $g_{\chi} = 2$, $g_* = 90$. }
  \end{center}
\end{figure}

\begin{figure}[h]
  \begin{center}
    \hspace*{-0.5cm} \includegraphics*[width=8cm]{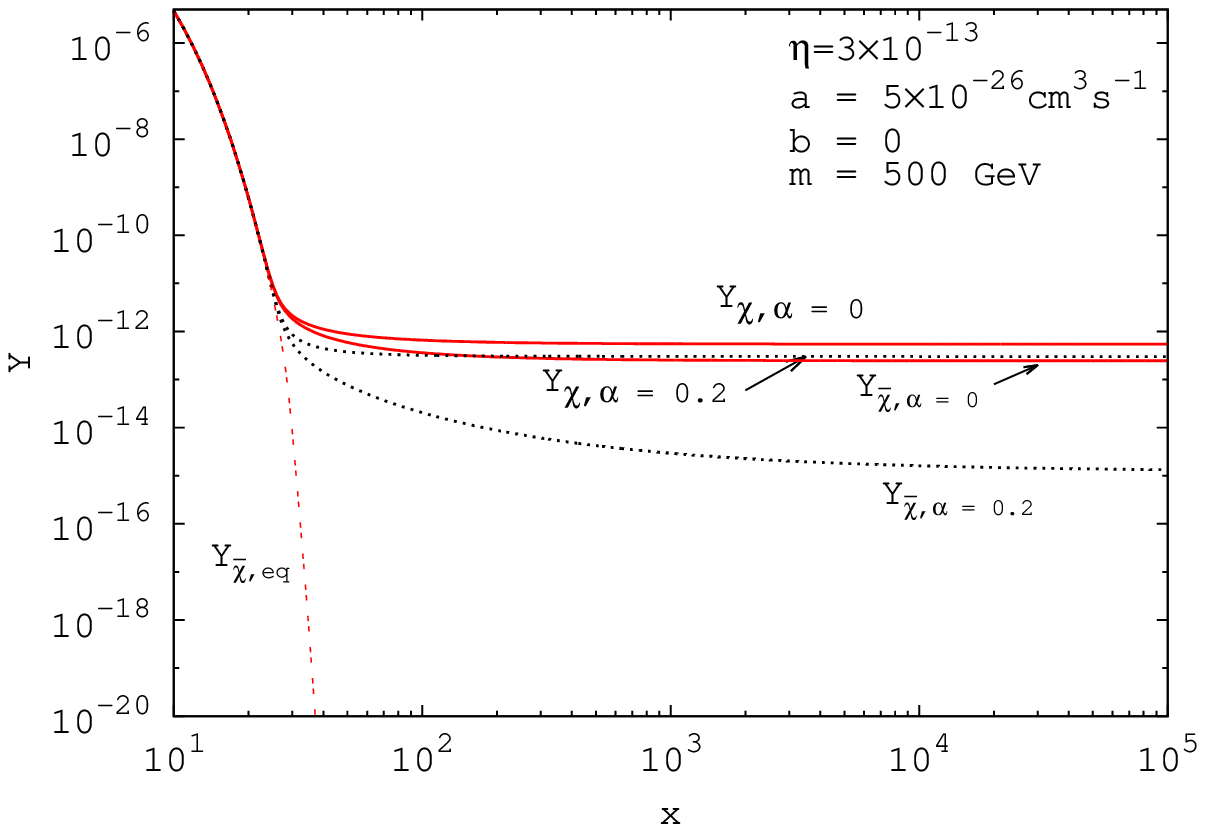}
    \put(-115,-12){(a)}
    \hspace*{-0.5cm} \includegraphics*[width=8cm]{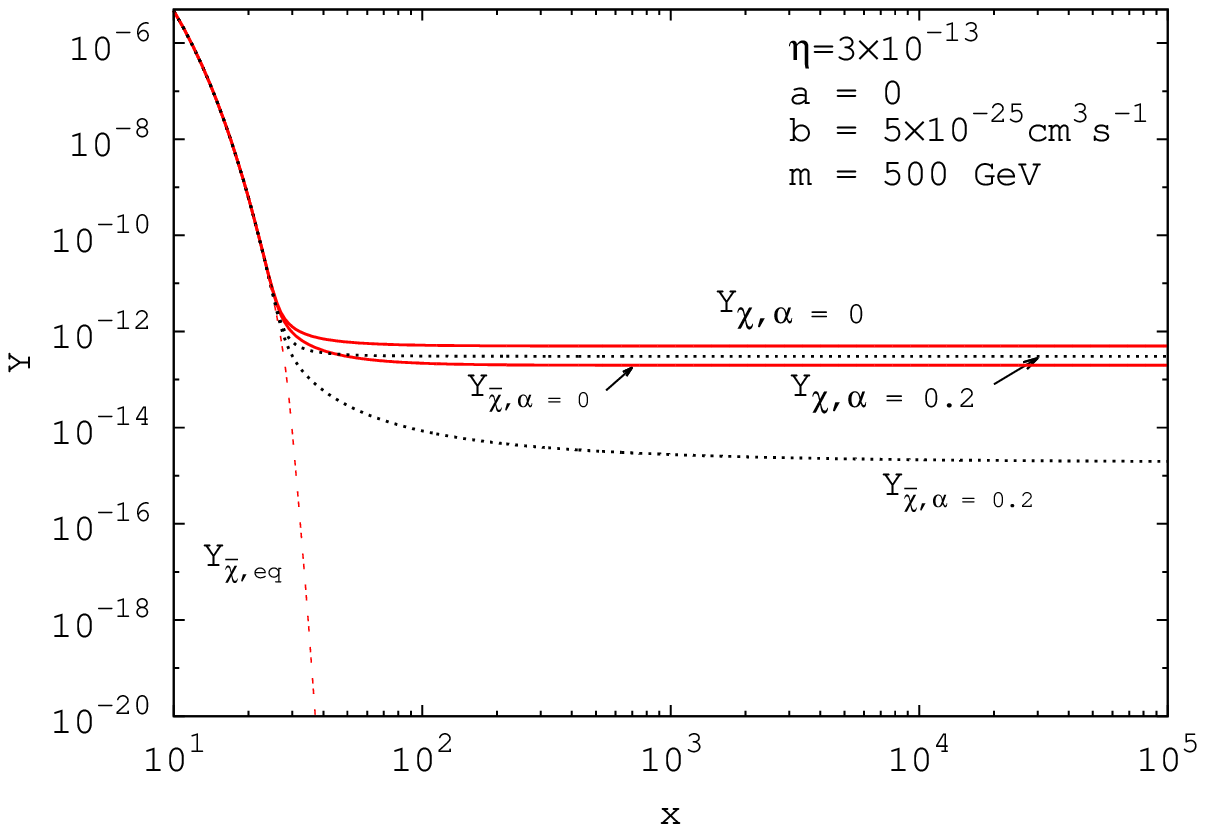}
    \put(-115,-12){(b)}
     \caption{\label{fig:k} \footnotesize Evolution of $Y$ for the particle
       and anti--particle as a function of $x$ for the case when the
       annihilation cross section is boosted by Sommerfeld enhancement and 
       without enhancement. Here $g_{\chi} = 2$, $g_* = 90$. }
  \end{center}
\end{figure}

Fig.\ref{fig:a} shows the evolution of abundances of Dark Matter
particle and anti--particle when the annihilation cross section is 
enhanced by Sommerfeld effect. It is plotted using the numerical solutions
of equations (\ref{eq:Yeta}), (\ref{eq:Ybareta}). In panel $(a)$, the 
thick (red) lines are for relic abundances $Y_{\chi}$ and $Y_{\bar\chi}$
for asymmetric Dark Matter 
particle and anti--particle without Sommerfeld effect. The dashed (blue) 
lines are for the case of Sommerfeld factor $\alpha=0.1$ and  
dotted (black) lines are for the case of Sommerfeld factor 
$\alpha = 0.2$. The double dotted (red) is for the
equilibrium value of anti--particle abundance. It is shown that 
deviations between the particle 
abundances of the case with Sommerfeld enhancement and without are very small 
for the case of $\alpha =0.1$ and $\alpha =0.2$. We found that the particle 
abundance is not affected  appreciably comparing to the anti--particle 
abundance. The impact of Sommerfeld enhancement on relic abundance 
of anti--particle is more significant when the Sommerfeld factor $\alpha$ is 
larger. The similar results is obtained for the case of $p$--wave
annihilation in plot $(b)$. The asymmetric Dark Matter 
decouples later due to the boosted annihilation rate comparing to the 
case without Sommerfeld enhancement, and hence the relic abundances for 
particle and anti--particle are decreased in principle. For 
$\alpha=0.1$ and $\alpha=0.2$ in Fig.\ref{fig:a}, the decreases of
anti--particle abundances are few 
orders less than $\eta$, 
then the particle abundances kept in the same order of $\eta$ due to 
the relation $Y_{\chi} - Y_{\bar\chi} = \eta$, because the anti--particle
abundance is too small to alter the particle abundance in 
Eq.(\ref{eq:eta}). This is the reason why the particle abundance is not changed 
sizably comparing to the anti--particle abundance.  For the smaller value of 
$\eta$, as in Fig.\ref{fig:k}, the decrease of asymmetric Dark Matter
particle abundance is obvious. 
\section{Analytical solutions}
We follow the method which is used in \cite{GSV,Iminniyaz:2011yp} to find the 
analytic solution, we first write the Boltzmann equation (\ref{eq:Ybareta}) 
in terms of $\Delta_{\bar\chi} = Y_{\bar\chi} - Y_{\bar\chi,{\rm eq}}$,
\begin{equation} \label{eq:deltabar}
\frac{d \Delta_{\bar\chi}}{dx} = - \frac{d Y_{\bar\chi,{\rm eq}}}{dx} -
\frac{{\lambda \langle \sigma v \rangle}_S}{x^2}~
     \left[\Delta_{\bar\chi}(\Delta_{\bar\chi} + 2 Y_{\bar\chi,{\rm eq}})
      + \eta \Delta_{\bar\chi}   \right]\, .
\end{equation}
For high temperature, $Y_{\bar\chi} \sim Y_{\bar\chi,{\rm eq}}$, therefore we
ignore $\Delta^2_{\bar\chi}$ and $d\Delta_{\bar\chi}/dx$, then
\begin{equation} \label{eq:bardelta_solu}
      \Delta_{\bar\chi} \simeq \frac{2 x^2 Y^2_{\rm eq}}   
      {\lambda {\langle \sigma v \rangle}_S\,(\eta^2 + 4 Y^2_{\rm eq})}\,,
 \end{equation}
here $Y_{\bar\chi,{\rm eq}} = - \eta/2 + \sqrt{\eta^2/4 + Y^2_{eq}}$, which is
obtained by solving the Boltzmann equation (\ref{eq:Ybareta}) in equilibrium
state. Eq.(\ref{eq:bardelta_solu}) is used to fix the freeze out
temperature $\bar{x}_F$ for $\bar{\chi}$. 

At late time, when the temperature is low, $x > \bar{x}_F$, the equilibrium 
value of relic abundance $Y_{\bar\chi,{\rm eq}}$ is negligible. Thus
after dropping the term which is related to $Y_{\bar\chi,{\rm eq}}$ in 
Eq.(\ref{eq:deltabar}), we have
\begin{equation} \label{eq:deltalate}
\frac{d \Delta_{\bar\chi}}{dx} = - \frac{\lambda {\langle \sigma v
\rangle}_S}{x^2 } \left( \Delta_{\bar\chi}^2 + \eta \Delta_{\bar\chi}
\right)\,,
\end{equation}
here we assume that 
$\Delta_{\bar\chi}(\bar{x}_F)\gg \Delta_{\bar\chi}(x_{\infty})$ and 
integrate Eq.(\ref{eq:deltalate}) from $\bar{x}_F$ to $\infty$, then
\begin{equation} \label{eq:barY_cross}
Y_{\bar\chi}(x_\infty) =  \eta\,\left\{ \,
  {\rm exp} \left[\, 1.32\, \eta \, m M_{\rm Pl}\,
     \sqrt{g_*} \, \int^{\infty}_{\bar{x}_F} 
     \frac{{\langle \sigma v \rangle}_S}{ x^2}dx\,\right] -1\, \right\}^{-1} \,,
\end{equation}
where 
\begin{eqnarray} \label{eq:integral}
     \int^{\infty}_{\bar{x}_F} 
     \frac{{\langle \sigma v \rangle}_S}{ x^2} dx &=& (a + \alpha^2\,b) 
        \left[ \frac{1}{\bar{x}_F} +  2 \alpha \sqrt{\frac{\pi}{\bar{x}_F}}
       +  \frac{\pi^2 \alpha^2}{6}~  {\rm ln} \left( 1 +  
        \frac{9 \alpha \sqrt{\pi \bar{x}_F} + 12 }
        {(9 - 2 \pi )\pi \alpha^2 \bar{x}_F }\right) \right.
      \nonumber \\      
       &+& \left.    \pi \alpha^2~  \frac{36 - 11 \pi}{\sqrt{3(117 - 32\pi)}}
        \left(  \frac{\pi}{2} - {\rm tan}^{-1}~ \frac{
       2(9 - 2\pi) \alpha \sqrt{\pi \bar{x}_F} +9}
        {\sqrt{3(117 - 32 \pi)}}  \right) \right]
     \nonumber \\
       & + & b\left[ \frac{3}{\bar{x}_F^2} + \frac{8  
   \sqrt{\pi} \alpha}{3\bar{x}_F^{3/2}} +  \frac{\pi^2 \alpha^2 }{3 \bar{x}_F} + 
        \frac{8 \pi^{5/2} \alpha^3 }
         {153 \sqrt{\bar{x}_F}} +
       \frac{(16 + 13\pi) \pi^3 \alpha^4}{459 \sqrt{ \pi/2 -1}}   
        \left( \frac{\pi}{2} - {\rm tan}^{-1}\frac{6 + 
     \pi^{3/2}  \alpha  \sqrt{\bar{x}_F}}{ 3\sqrt{2(\pi -2)}}  \right) \right.
         \nonumber \\
            &  - &  \left.
         \frac{(16 + 17\pi) \pi^3 \alpha^4 }{918}~{\rm ln}
        \left( 1 + \frac{12}{\pi \alpha \sqrt{\pi \bar{x}_F}} +  
  \frac{18}{\pi^2 \alpha^2 \bar{x}_F}  \right) \right]\,  .                     
\end{eqnarray} 
The relic abundance for $\chi$ particle is obtained by using Eq.(\ref{eq:eta}), 
\begin{equation}\label{eq:Y_fin}
      Y_{\chi}(x_\infty) = \eta\,
 \left\{\, 1 - \exp \left[\, - 1.32\,  \eta\, m M_{\rm Pl}\,
 \sqrt{g_*} \,   \int^{\infty}_{x_F} 
     \frac{{\langle \sigma v \rangle}_S}{ x^2}dx \,   \right]\, \right\}^{-1}\,,
\end{equation}
where $x_F$ is the freeze out temperature for $\chi$. 
Eqs.(\ref{eq:barY_cross}) and (\ref{eq:Y_fin}) are only consistent with
constraint (\ref{eq:eta}) if $x_F = \bar{x}_F$. The total final Dark 
Matter relic density is  
\begin{eqnarray} \label{eq:omega}
 \Omega_{\rm DM}  h^2 & = & 2.76 \times 10^8\, \left[ Y_{\chi}(x_\infty) + 
                       Y_{\bar\chi}(x_\infty) \right]\,m,
\end{eqnarray}
where we used $\Omega_{\chi} = \rho_{\chi}/\rho_c$ with 
$\rho_{\chi}=n_{\chi} m = s_0 Y_{\chi}  $ and 
$\rho_c = 3 H^2_0 M^2_{\rm Pl}$, here $s_0 \simeq 2900$ cm$^{-3}$ is the 
present entropy density, and $H_0$ is the Hubble constant. We use the
equality, $\xi Y_{\bar\chi,{\rm eq}}( \bar{x}_F) = \Delta_{\bar\chi}(\bar{x}_F) $,
to fix the freezing out temperature, here $\xi$ is a constant, 
usually we take $\xi = \sqrt{2} -1$ \cite{standard-cos}. We found
the analytic result matches with the numerical result within the accuracy 
of $10\%$.

\section{Effects of kinetic decoupling on the relic abundance of asymmetric
 Dark Matter}
The effects of Sommerfeld enhancement on the relic density of 
asymmetric Dark Matter is analyzed in the previous section. It was assumed the
temperatures of annihilating asymmetric Dark Matter particles and
anti--particles track the background radiation temperature $T$ when the
annihilating asymmetric Dark Matter particles and anti--particles remains in
chemical and kinetic equilibrium with the radiation background. During
radiation dominated era, the temperature of radiation scales as 
$T \propto 1/R$, with $R$ being the scale factor of the universe. 
Asymmetric Dark Matter particle and anti--particle are still in kinetic 
equilibrium after
droping out of chemical equlibrium. At some point $T_k$, asymmetric
Dark Matter particles and anti--particles decouple from kinetic equilibirum
and the temperature of asymmetric Dark Matter scales as 
$T_{\chi,\bar\chi} \propto 1/R^2$ 
\cite{Bringmann:2006mu,Bringmann:2009vf,Chen:2001jz,Hofmann:2001bi}. The 
determination of precise value of the kinetic
decoupling temperature $T_k$ depends on the models. In supersymmetric 
models discussed in \cite{Bringmann:2009vf}, 
$T_k \approx (10^{-3} - 10^{-1}) T_F$. 
In this work, we take $T_k/T_F$ as a free 
parameter for the generality with the constraint $T_k < T_F$. 
Then the relation between the temperatures of asymmetric Dark Matter 
$T_{\chi,\bar\chi}$ and the radiation temperature $T$ is 
\cite{Bringmann:2006mu,Bringmann:2009vf}
\begin{equation}\label{eq:tempera}
T_{\chi,\bar\chi} = \frac{T^2}{T_k}.
\end{equation}

This change will affect the thermal average of the annihilation cross section
between the asymmetric Dark Matter particle and anti--particle.
For the case of $s-$wave annihilation, the cross section is independent of 
$T_{\chi,\bar\chi}$, therefore, kinetic decoupling has no effect on the relic 
density of asymmetric Dark Matter in this case. For the $p-$wave 
annihilation or Sommerfeld enhanced $s-$ and $p-$wave annihilations, there 
are temperature dependencies of the annihilation cross section, then the relic 
density is affected by kinetic decoupling. After kinetic 
decoupling, thermal average of $p-$wave annihilation cross section becomes 
${\langle \sigma v\rangle}_p = 6b\, x_k/x^2 $. The 
Boltzmann equations of asymmetric Dark Matter anti--particle for $p-$wave 
annihilation before and after kinetic decoupling are
\begin{equation} \label{eq:boltzmann_Ypw}
\frac{d Y_{\bar\chi}}{dx} =
      -  1.32\,m M_{\rm Pl}\, \sqrt{g_*} (6b\, x^{-3})\,
     (Y_{\bar\chi}^2 + \eta Y_{\bar\chi}  - Y^2_{\rm eq})\,,
\end{equation}
\begin{equation} \label{eq:boltzmann_Ypwk}
\frac{d Y_{\bar\chi}}{dx} =
      -  1.32\,m M_{\rm Pl}\, \sqrt{g_*} (6b\, x_k x^{-4})\,
     (Y_{\bar\chi}^2 + \eta Y_{\bar\chi}  - Y^2_{\rm eq})\,,
\end{equation}

The effects of kinetic decoupling on the final relic density of asymmetric
Dark Matter for $p-$wave annihilation is estimated by integrating Boltzmann 
equation (\ref{eq:boltzmann_Ypw}) from 
$\bar{x}_F$ to $x_k$ and equation (\ref{eq:boltzmann_Ypwk}) from 
$x_k$ to $\infty$. 
When there is kinetic decoupling, we obtain the relic abundance 
for asymmetric Dark Matter anti--particle for $p-$wave annihilation as 
\begin{equation} \label{eq:barY_crossp}
Y_{\bar\chi}(x_{\infty}) =  \eta\,\left\{ \,
  {\rm exp} \left[\, 1.32\, \eta \, m M_{\rm Pl}\,
     \sqrt{g_*} \, \left( \int^{x_k}_{\bar{x}_F} 
     \frac{6b }{ x^3}\,dx +
     \int^{\infty}_{x_k} 
     \frac{6bx_k}{ x^4}\,dx \right)
      \right] -1\, \right\}^{-1} \,.
\end{equation}

\begin{figure}[h!]
  \begin{center}
    \hspace*{-0.5cm} \includegraphics*[width=10cm]{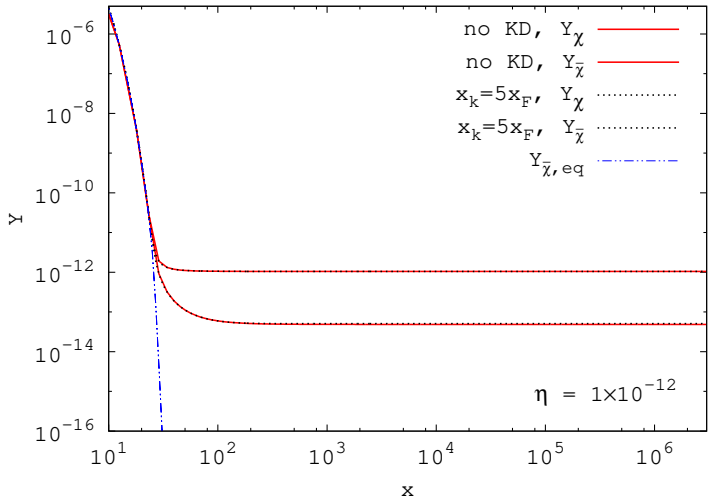}
    \put(-115,-12){}
     \caption{\label{fig:b} 
     \footnotesize The effect of kinetic decoupling on the evolution of $Y$ 
     for the particle and anti--particle as a function of $x$ for $p-$wave 
     annihilation cross section. Here $g_{\chi} = 2$, $g_* = 90$,
     $m=500$ GeV, $x_F =25$, $a=0$, 
     $b = 5 \times 10^{-25}$ ${\rm cm}^3 \,{\rm s}^{-1}$,}
  \end{center}
\end{figure}
In Fig.\ref{fig:b}, we plot the relic abudances of asymmetric Dark Matter 
particle $Y_{\chi}$ and 
anti--particle $Y_{\bar\chi}$ as a function of the inverse--scaled temperature
$x$ for $p-$wave annihilation cross section when the kinetic 
decoupling temperature $x_k = 5 x_F$, here $\alpha =0$, $a=0$,
$b = 5 \times 10^{-25}$ ${\rm cm}^3 \,{\rm s}^{-1}$, $\eta = 1 \times 10^{-12}$ 
and $m=500$ GeV. The effects of kinetic 
decoupling on the asymmetric Dark Matter particle abundance $Y_{\chi}$ and 
anti--particle abundance $Y_{\bar\chi}$ are negligible when kinetic decoupling 
temperature $x_k = 5 x_F$. The Dark Matter particle 
abundance is almost not changed after kinetic decoupling. The difference 
between the anti--particle abundance before 
and after kinetic decoupling is by a factor of 1.  Because we are discussing 
the case where kinetic decoupling occured after the asymmetric Dark Matter
particles and anti--particles decoupled from chemical equilibrium, again here 
we assume the kinetic decoupling occurred at the point which is 5 times 
of inverse--scaled chemical decoupling temperature, therfore, the effect is 
negligible in this case. It
may have significant effects if the kinetic decoupling occurs earlier. In that 
case, one must solve the coupled Boltzmann equations which we didn't consider 
in our work for simplicity \cite{Binder:2017rgn}.

The effect of kinetic decoupling is more noticeable for the 
case of Sommerfeld enhanced $s-$wave and $p-$wave annihilations.
With the kinetic decoupling, the Sommerfeld enhanced annihilation cross 
sections become 
\begin{eqnarray}\label{eq:Somcrosskinet}
     {\langle \sigma v \rangle}_{S_k} & \simeq & 
      \frac{x^3}{2 \sqrt{\pi x^3_k}}\,\int^{\infty}_0 dv\,e^{-\frac{x^2}{4x_k} v^2}
 \Bigg\{ a\,v^2 \, 
          \frac{2 \pi \alpha/v}{1- e^{- 2 \pi \alpha/v}} +  
          b\,v^4 \,  
          \left[ 1 + (\frac{\alpha}{v})^2 \right]
          \frac{2 \pi \alpha/v}{1- e^{- 2 \pi \alpha/v}}\Bigg\}\,  . 
\end{eqnarray} 
Then the Boltzmann equation (\ref{eq:Ybareta}) of asymmetric Dark Matter
anti--particle for Sommerfeld enhanced $s-$ and $p-$wave annihilation cross
sections is
\begin{equation} \label{eq:boltzmann_Yk}
\frac{d Y_{\bar\chi}}{dx} =
      - { 1.32\,m M_{\rm Pl}\, \sqrt{g_*}
      \langle \sigma v \rangle}_{S_k}\,x^{-2}\,
     (Y_{\bar\chi}^2 + \eta Y_{\bar\chi}  - Y^2_{\rm eq})\,,
\end{equation}
\begin{figure}[t!]
  \begin{center}
    \hspace*{-0.5cm} \includegraphics*[width=8cm]{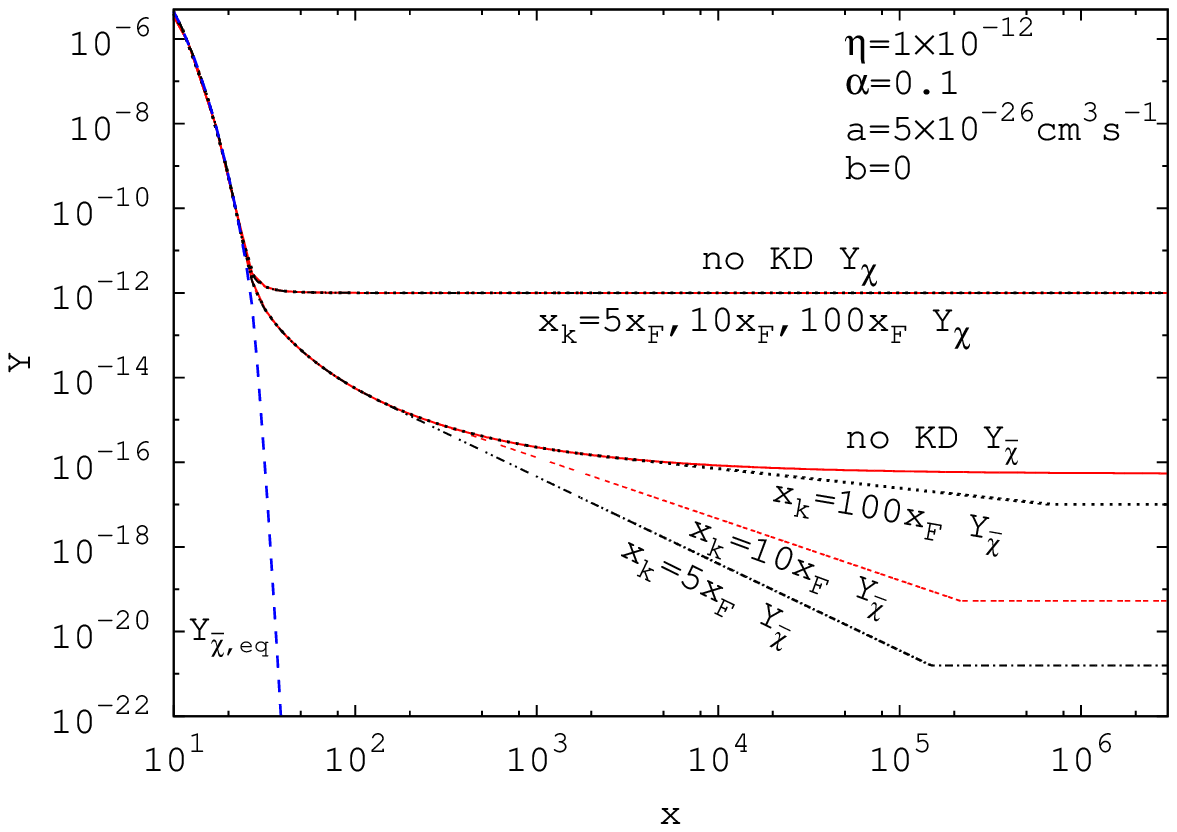}
    \put(-115,-12){(a)}
    \hspace*{-0.5cm} \includegraphics*[width=8cm]{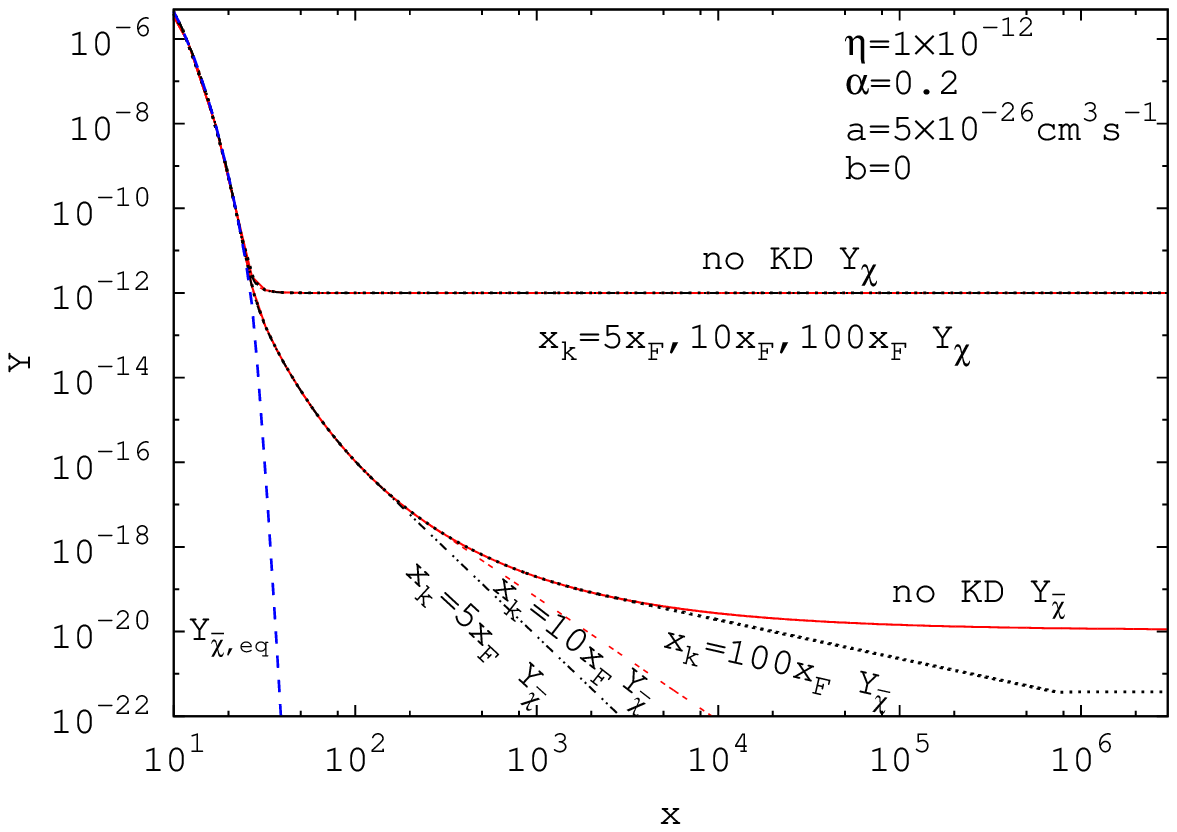}
    \put(-115,-12){(b)}
     \vspace{0.5cm}
     \hspace*{-0.5cm} \includegraphics*[width=8cm]{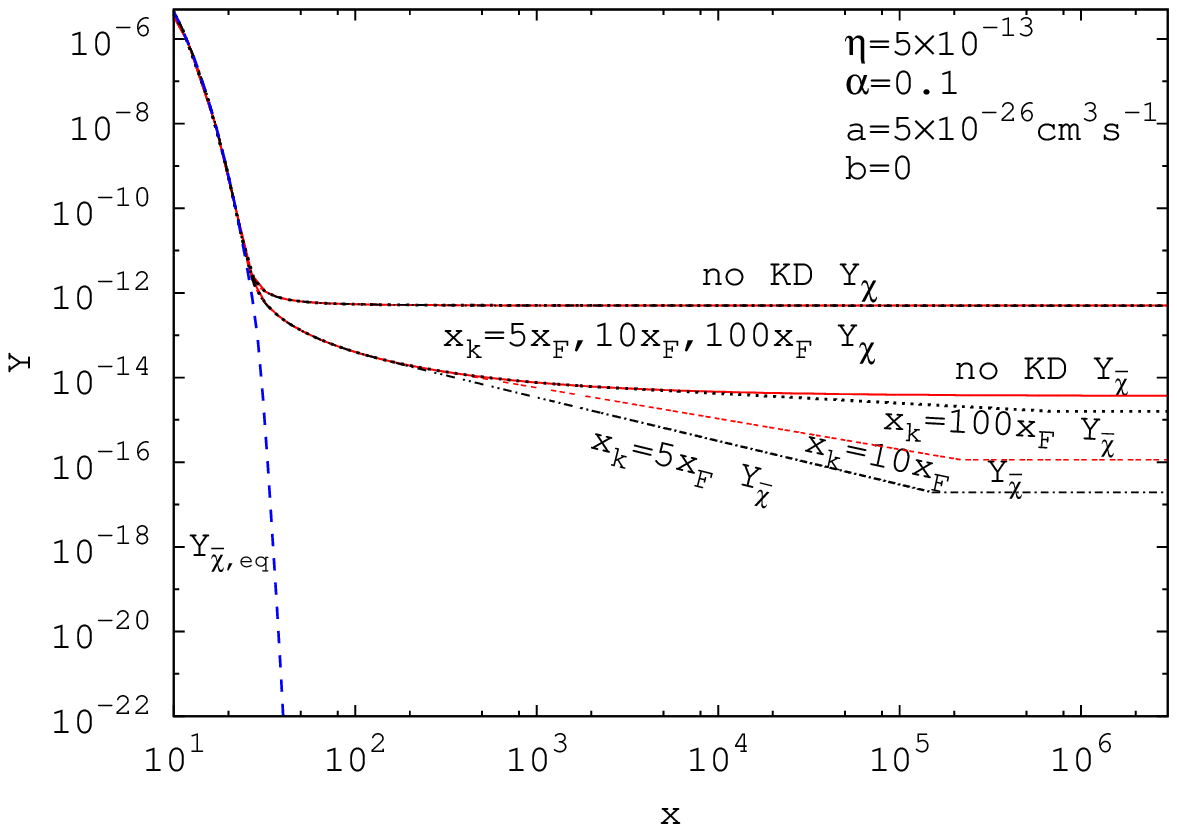}
    \put(-115,-12){(c)}
    \hspace*{-0.5cm} \includegraphics*[width=8cm]{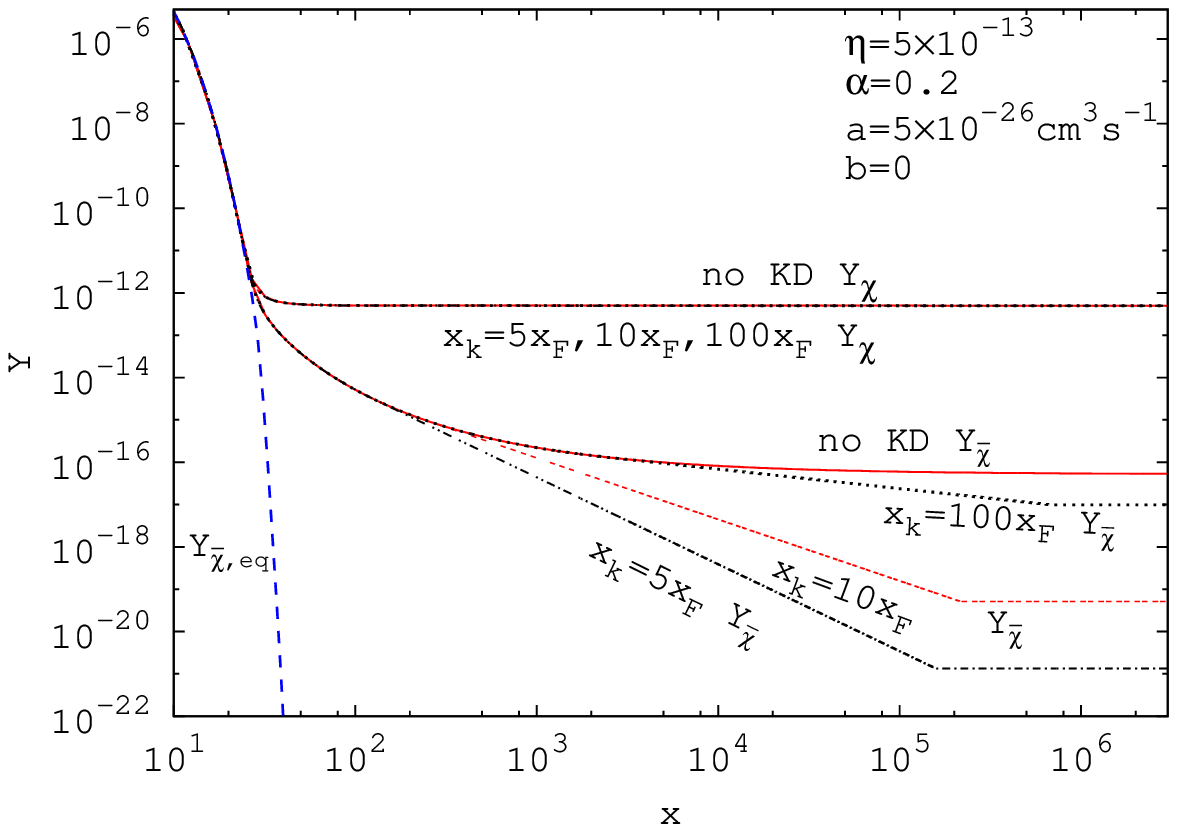}
    \put(-115,-12){(d)}
    \caption{\label{fig:c} 
     \footnotesize The effects of kinetic decoupling on the evolution of $Y$ 
     for asymmetric Dark Matter particle and anti--particle as a function 
     of $x$ for Sommerfeld enhanced $s-$wave annihilation cross section for 
     different asymmetry factors and coupling strengths. 
     Here $g_{\chi} = 2$, $g_* = 90$, $m=500$ GeV, $x_F =25$.}
   \end{center}
\end{figure}
Fig.\ref{fig:c} shows the 
evolution of $Y_{\chi}$ and $Y_{\bar\chi}$ as a function of $x$ for $s-$wave 
annihilation cross section when $\alpha = 0.1$ in panels (a), (c) and 
$\alpha = 0.2$ in panels (b), (d). 
Here the asymmetry factor $\eta = 1 \times 10^{-12}$ in panels (a), (b);
 $\eta = 5 \times 10^{-13}$ in (c), (d) and $m=500$ GeV,
$a = 5 \times 10^{-26}$ ${\rm cm}^3 \,{\rm s}^{-1}$, $b = 0$. We 
plot the figure by using the numerical solution of Eq.(\ref{eq:Ybareta}) from
the range of $\bar{x}_F$ to $x_k$ and Eq.(\ref{eq:boltzmann_Yk}) from 
$x_k$ to quite large value of $x$, here we take $x= 3 \times 10^6$. We 
found the
asymmetric Dark Matter particle abundances for different kinetic decoupling 
temperatures are
almost same with the case that there is no kinetic decoupling. On the other 
hand, after kinetic decoupling, the relic abundances for 
anti--particle are decreased continuously until the annihilation becomes
inefficient. If we replace $x$ with $x_{\chi} = x^2/x_k $ in the
analytic result of the $s-$wave Sommerfeld factor in Eq.(\ref{eq:approxs}),
the Sommerfeld factor $\propto x$ for sufficiently large $x$. After the 
integration of equation 
({\ref{eq:boltzmann_Yk}}), for large $x$, the
anti--particle abundance for $s-$wave annihilation cross section scales as 
$Y_{\bar\chi} \propto \eta/x^c$, where 
$c \propto 1.32 \eta\, m\, M_{\rm Pl}\,\sqrt{g_*}\,\alpha\,a$, which is 
constant. It 
matches with the numerical result. However, this decrease will eventually be
stopped by one of the following three effects \cite{vandenAarssen:2012ag}. 
One is that 
the Sommerfeld enhancement is saturated at low velocity, it works for the 
massive mediator case. Second is that the onset of matter domination. The 
last one is the onset of structure formation which finally eliminates the 
Sommerfeld effect. We use $x_{\rm cut}$ to express the point at 
which the Sommerfeld effect is eliminated. In plot $(a)$, 
the relic abundances become constant around $x_{\rm cut}=1.5 \times 10^5$ for 
$x_k =5x_F$, $2.2 \times 10^5$ for $x_k =10x_F$ and $6.9 \times 10^5$ for 
$x_k=100x_F$. We obtained these points from the numerical data. The asymmetric
Dark Matter annihilation rate is insignificant from that points and
$Y_{\bar\chi}$ becomes stable. The inverse--scaled 
temperature at which the
annihilations become inefficient is important for the correct determination of 
the relic density of asymmetric Dark Matter.
The decrease of abundance of Dark Matter anti--particle is larger when the
decoupling temperature is more close to the chemical freezing out point $x_F$. 
The reduction is also more sizable for larger $\alpha$. 
For the smaller asymmetry factor $\eta = 5 \times 10^{-13} $, the decreases of 
Dark Matter anti--particle abundances are less than the case of 
$\eta = 1 \times 10^{-12}$ which are shown in panels $(c)$ and $(d)$. 
%
%
\begin{figure}[h]
  \begin{center}
    \hspace*{-0.5cm} \includegraphics*[width=8cm]{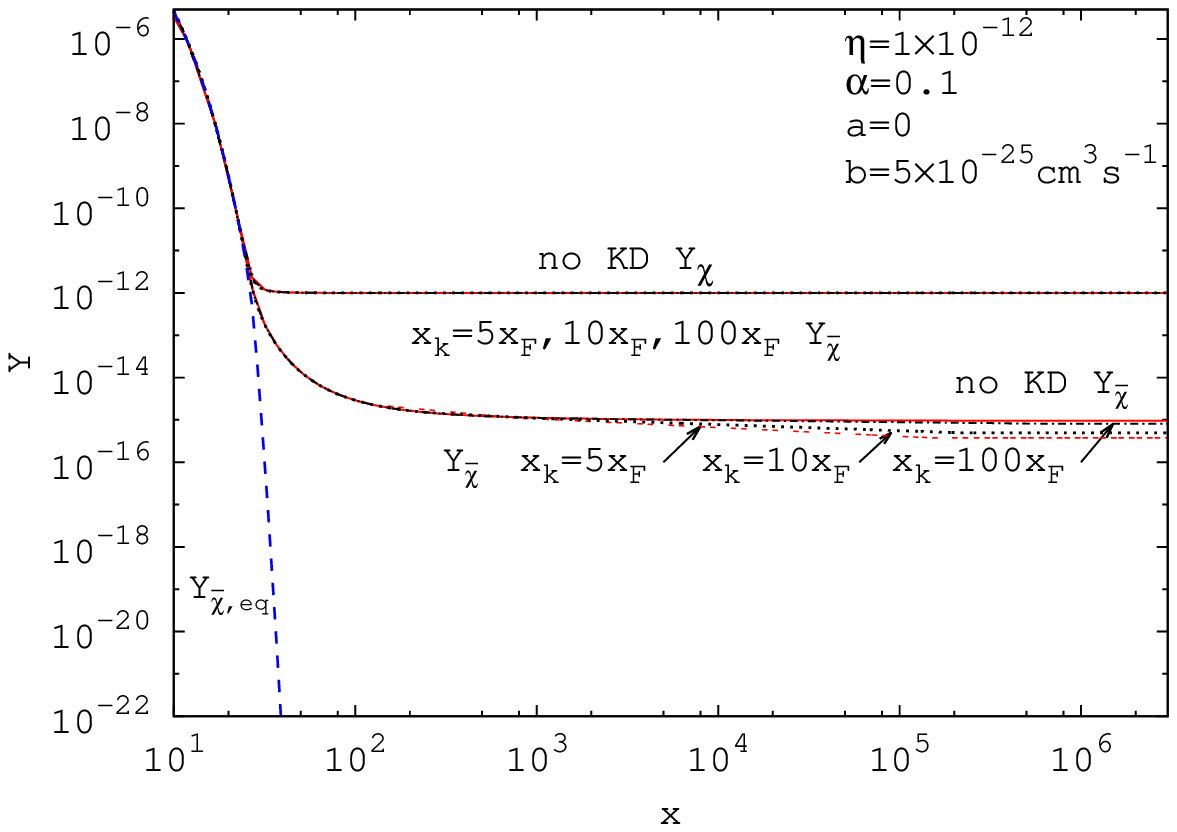}
    \put(-115,-12){(a)}
    \hspace*{-0.5cm} \includegraphics*[width=8cm]{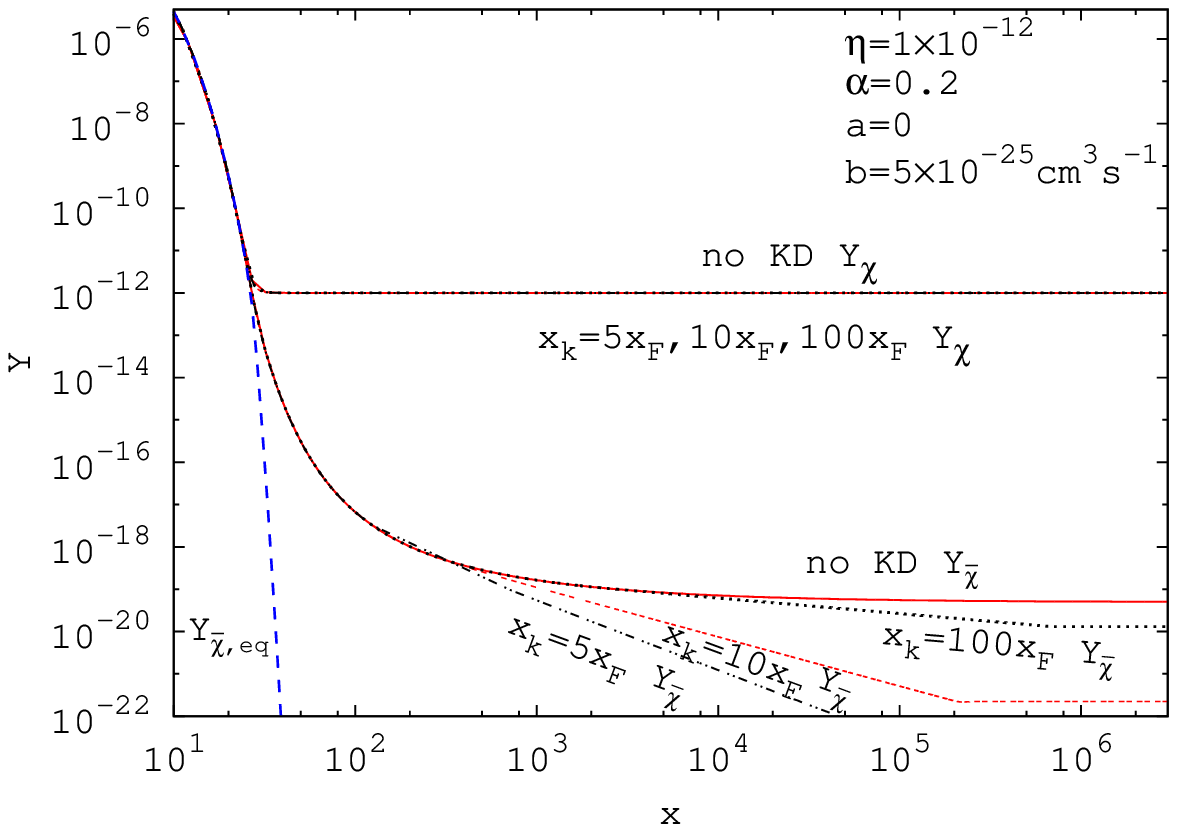}
    \put(-115,-12){(b)}
     \caption{\label{fig:l} \footnotesize The effects of kinetic decoupling on 
     the evolution of $Y$ for the particle and anti--particle as a function 
     of $x$ for Sommerfeld enhanced $p-$wave annihilation cross section.
     Here $g_{\chi} = 2$, $g_* = 90$, $m=500$ GeV, $x_F =25$. }
  \end{center}
\end{figure}

The cases of $\alpha = 0.1$ and $\alpha = 0.2$ for Sommerfeld 
enhanced $p-$wave 
annihilation cross section are plotted in Fig.\ref{fig:l} for kinetic 
decoupling temperatures $x_k = 5 x_F ,10x_F, 100x_F$. Here 
$\eta = 1 \times 10^{-12}$, $m=500$ GeV, $a = 0$, $b = 5 \times 10^{-25}$ 
${\rm cm}^3 \,{\rm s}^{-1}$. Similar analysis with the $s-$wave annihilation 
can be done for the case of $p-$wave annihilation. 
The abundances for asymmetric Dark Matter particles are nearly not changed for
different kinetic decoupling temperatures. 
For asymmetric Dark Matter anti--particle, the decrease of abundance is very 
small for $\alpha = 0.1$ comparing to the case $\alpha = 0.2$. On the other 
hand, the decrease is larger for smaller 
inverse--scaled kinetic decoupling temperature $x_k =5x_F$ in the case when 
$\alpha = 0.2$. In panel $(b)$, annihilations become 
insignificant at the point $x_{\rm cut} = 4.1 \times 10^5$ for 
$x_k = 5x_F$, $2.6\times 10^5$ for $x_k =10x_F$ and $6.9\times 10^5$
for $x_k =100x_F$.

The final relic abundance for asymmetric Dark Matter 
anti--particle for Sommerfeld enhanced $s$, $p-$wave annihilations is 
obtained by integrating Boltzmann equation (\ref{eq:Ybareta}) from 
$\bar{x}_F$ to $x_k$ and equation (\ref{eq:boltzmann_Yk}) from 
$x_k$ to $x_{\rm cut}$, then
\begin{equation} \label{eq:barY_crossn}
Y_{\bar\chi}(x_{\rm cut}) =  \eta\,\left\{ \,
  {\rm exp} \left[\, 1.32\, \eta \, m M_{\rm Pl}\,
     \sqrt{g_*} \, \left( \int^{x_k}_{\bar{x}_F} 
     \frac{{\langle \sigma v \rangle}_S}{ x^2}\,dx +
     \int^{x_{\rm cut}}_{x_k} 
     \frac{{\langle \sigma v \rangle}_{S_k}}{ x^2}\,dx \right)
      \right] -1\, \right\}^{-1} \,.
\end{equation}  
\section{Constraints}
Dark Matter relic density provided by Planck data \cite{Ade:2015xua} is  
\begin{eqnarray} \label{eq:pldata}
  \Omega_{\rm DM} h^2 = 0.1199 \pm 0.0022\, .
\end{eqnarray}
\begin{figure}[t!]
  \begin{center}
    \hspace*{-0.5cm} \includegraphics*[width=8cm]{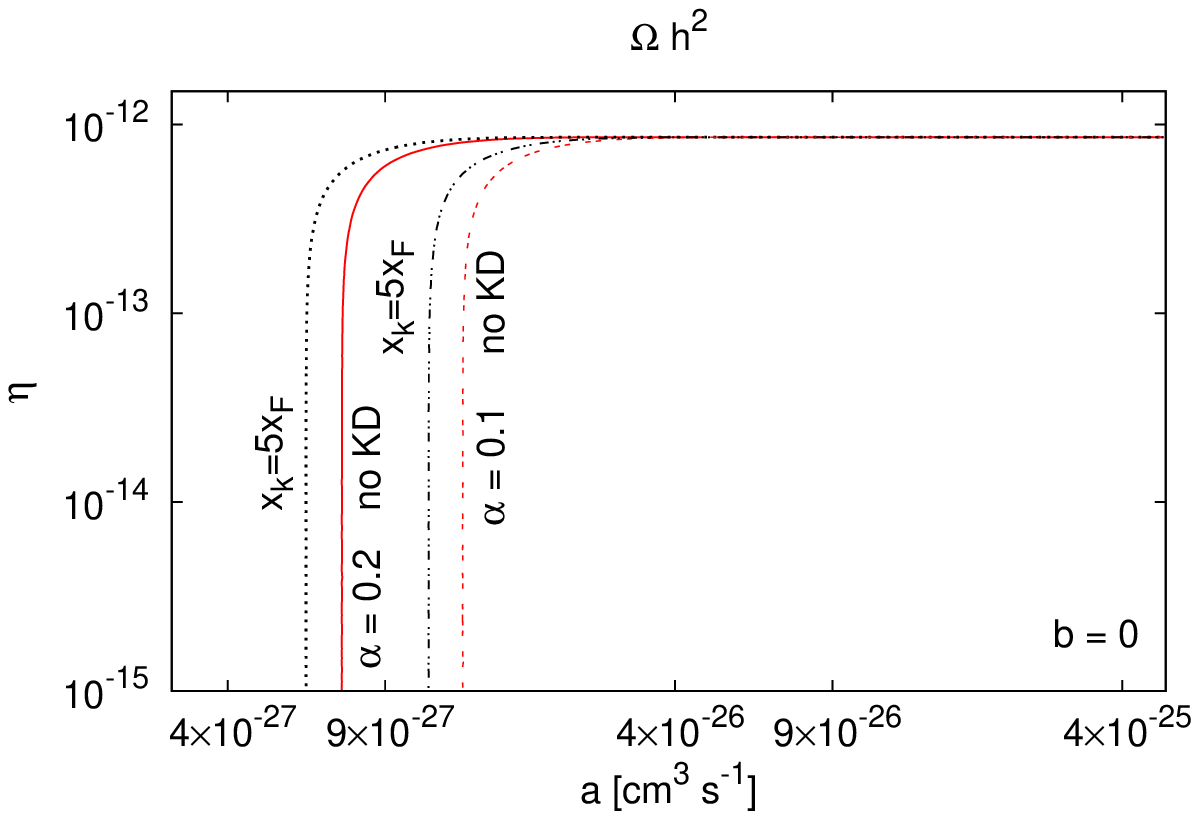}
    \put(-115,-12){(a)}
    \hspace*{-0.5cm} \includegraphics*[width=8cm]{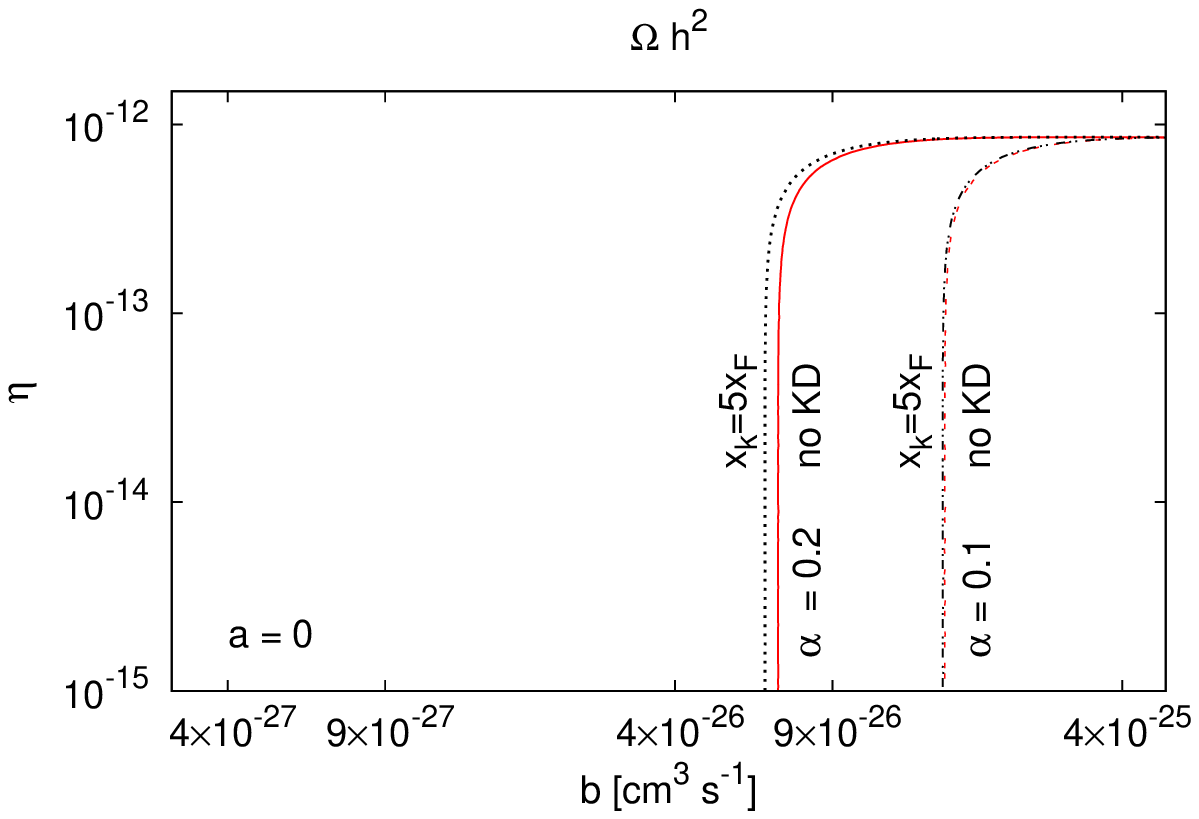}
    \put(-115,-12){(b)}
     \caption{\label{fig:e} 
     \footnotesize The contour plots of $s-$ ($b=0$) and $p-$wave ($a=0$) 
     annihilation cross sections and the asymmetry factor $\eta$ when 
     $\Omega_{\rm DM} h^2 = 0.1199$. 
     Here $g_{\chi} = 2$, $g_* = 90$, $m=500$ GeV, $x_F =25$. }
   \end{center}
\end{figure}
Fig.\ref{fig:e} shows the contour plots of $s-$ (panel $(a)$) and 
$p-$wave (panel $(b)$) 
annihilation cross sections and asymmetry factor $\eta$ when 
$\Omega_{\rm DM} h^2 = 0.1199$. The 
loosely dashed (red) line is for the case of Sommerfeld enhancement without 
kinetic decoupling and dash dotted (black) line is for the case of 
kinetic decoupling when $\alpha = 0.1$, here the inverse--scaled kinetic 
decoupling temperature is $x_k=5x_F$. The thick (red) line is for the case 
when there is no kinetic decoupling and the dotted (black) line is for 
inverse--scaled kinetic decoupling temperature 
$x_k =5x_F$ when $\alpha = 0.2$. We found the required annihilation cross
section with kinetic decoupling is smaller than the case of without
kinetic decoupling, i.e. when $\alpha = 0.2$ and 
$\eta = 1.0\times 10^{-15}$ in panel (a), the
required cross section is 
$a = 5.99 \times 10^{-27}$ ${\rm cm}^3 \,{\rm s}^{-1}$ for the case of kinetic
decoupling and $a = 7.20 \times 10^{-27}$ ${\rm cm}^3 \,{\rm s}^{-1}$ for the
case of no kinetic decoupling. The reason is that the relic density is decreased
continuously after kinetic decoupling until the annihilation becomes 
inefficient. As a result there is less relic density for the case of kinetic
decoupling comparing to the case of without kinetic decoupling. In order to
satisfy the observed range of Dark Matter relic density, when there is kinetic
decoupling, the annihilation cross section should be smaller than the case of 
without kinetic decoupling. On the other hand, the required annihilation cross 
section for $\alpha=0.2$ is two times smaller than the case of 
$\alpha = 0.1$. We can see the reason from Fig.\ref{fig:c}, the decrease of
asymmetric Dark Matter anti--particle abundance is larger for larger coupling 
strength $\alpha$. Similar analysis can be done for the
case of $p-$wave annihilation cross section in panel (b) in Fig.\ref{fig:e},
i.e. for $\alpha = 0.2$ and $\eta = 1.0 \times 10^{-15}$, the required 
cross section is $b = 6.35 \times 10^{-26}$ ${\rm cm}^3 \,{\rm s}^{-1}$ for 
the case of kinetic decoupling and 
$b = 6.80 \times 10^{-26}$ ${\rm cm}^3 \,{\rm s}^{-1}$ for the
case of no kinetic decoupling. The difference of the required cross section 
between the kinetic decoupling and no kinetic decoupling is very small for 
$\alpha = 0.1$ for $p-$wave 
annihilation. We can find the reason from panel (a) and (b) of 
Fig.\ref{fig:l}. After kinetic decoupling, asymmetric Dark matter particle 
abundance is almost same for $\alpha = 0.1$ and $\alpha = 0.2$. The decrease 
of anti--particle abundance for $\alpha = 0.1$ is very small
in panel (a) compared to the case of $\alpha = 0.2$ in panel (b).

\section{Summary and conclusions}

We investigated the relic density of asymmetric Dark Matter which is coupled 
to the light force mediator. When the mediator is light enough, the 
interaction between the asymmetric Dark Matter particle and anti--particle 
is emerged as long--range interaction which distorts the wavefunction of 
two incoming asymmetric Dark Matter particle and anti--particle . It is indeed 
the Sommerfeld effect which enhances the annihilation rate of asymmetric 
Dark Matter at low velocity. The relic density of asymmetric Dark Matter is 
explored when the annihilation cross section is boosted by the Sommerfeld 
effect. First, we found the thermal average of Sommerfeld enhanced 
annihilation cross section. Then we derive the analytic formulae for relic 
abundances of asymmetric Dark Matter particle and anti--particle. 
We found the abundance for asymmetric Dark Matter particle is not 
affected too much. On the other hand, the decrease
of the relic abundance of asymmetric Dark Matter anti-particle is more
obvious than the particle due to the Sommerfeld enhancement. The size of  
decrease depends on the Sommerfeld factor $\alpha$. For larger $\alpha$, 
there is sizable decrease of the relic abundance. 

Then, we discuss the effects of kinetic decoupling on the relic abundances 
of asymmetric Dark Matter particle and anti--particle when the annihilation 
cross section of asymmetric Dark Matter is changed by the Sommerfeld
effect. After chemical decoupling, the asymmetric Dark Matter particles 
and anti--particles continue to keep in kinetic equilibrium. When the 
scattering rate falls below the expansion rate of the universe, asymmetric 
Dark Matter
particles and anti--particles decouple from kinetic equilibrium. The
temperatures of asymmetric Dark Matter are different before and after kinetic
decoupling. This leaves its imprint on the relic density of asymmetric
Dark Matter particle and anti--particle. There is no effect on the
$s-$wave annihilation while the impact is almost negligible for $p-$wave
annihilation when there is no Sommerfeld enhancement. On the other hand, when 
the annihilation cross section is
increased by the Sommerfeld enhancement, there are quite significant effects on
relic density of asymmetric Dark Matter both for $s-$ and $p-$wave 
annihilation cross sections.  

In our work, we assumed that kinetic decoupling occurred after the chemical 
decoupling. The kinetic decoupling point is at least 5 times of the 
inverse--scaled freezing out temperature. We found the 
decrease is negligible for the abundance of asymmetric Dark Matter particle. 
The asymmetric Dark Matter anti--particle abundance is continuously decreased 
after the kinetic decoupling until the annihilations become insignificant.
The magnitude of decrease depends on the size of kinetic decoupling 
temperature, the coupling strength $\alpha$ and asymmetry factor $\eta$. The 
decrease is larger when the kinetic decoupling temperature is more close to the 
freezing out point. The reduction of anti--particle abundance is more
sizable for larger $\alpha$ and also for larger asymmetry factor $\eta$. 

Finally, we used Planck data and found the constraints on  
annihilation cross section and asymmetry factor when there is kinetic 
decoupling and no kinetic decoupling. Our results show the required cross 
section for Dark Matter 
should be smaller than the case of without kinetic decoupling in order to 
fall in the observation range of Dark Matter relic density. It is
because there is less relic density of asymmetric Dark Matter due to the
kinetic decoupling. The result is important for determining the relic 
abundance of asymmetric Dark Matter when the Sommerfeld effect plays the 
role in low velocity.  
Sommerfeld effects imply the indirect detection signals from
the annihilations of asymmetric Dark Matter anti--particle is significant.
This provides us the possibility to probe the asymmetric Dark Matter with
observations of the CMB (Cosmic Microwave Background), the Milky way and 
Dwarf galaxies. 


\section*{Acknowledgments}

The work is supported by the National Natural Science Foundation of China
(11765021).

\end{document}